\documentclass[a4paper,11pt]{article}
\usepackage{pos}
\usepackage{subfig}
\usepackage{diagbox}
\usepackage{makecell, multirow}
\allowdisplaybreaks

\newcommand{\fig}[1]{Fig.~\ref{#1}}

\newcommand{\refcite}[1]{Ref.~\cite{#1}}

\title{Exploring Single-Flavor Dibaryons: A lattice perspective }

\author*[1]{Navdeep Singh Dhindsa}
\author[2]{Nilmani Mathur}
\author[1,3]{M.~Padmanath}

\affiliation[1]{The Institute of Mathematical Sciences,  CIT Campus, Chennai, 600113, India}
\affiliation[2]{Department of Theoretical Physics, Tata Institute of Fundamental Research, Homi Bhabha Road, Colaba, Mumbai 400005, India}
\affiliation[3]{Homi Bhabha National Institute, Training School Complex, Anushaktinagar, Mumbai 400094, India}

\emailAdd{navdeep.s.dhindsa@gmail.com}

\FullConference{The 41st International Symposium on Lattice Field Theory (LATTICE2024)\\
 28 July - 3 August 2024\\
Liverpool, UK\\}

\abstract{
We present a lattice calculation of dibaryons composed of single-flavor quarks with either charm or strange quark mass. We utilize a set of lattice QCD ensembles with $N_f=2+1+1$ dynamical HISQ fields, two spatial volumes, and four different lattice spacings generated by the MILC collaboration. By using an overlap action for the valence quark propagators, we calculate the ground state energies of dibaryons in $S = 0$ and $S = 2$ spin channels. By analyzing the energy difference of the ground state of the dibaryon with respect to the relevant threshold, we provide insights into the interactions involved in different spin channels at the charm and the strange quark masses.}

\begin{document}
\maketitle

\section{\label{sec:intro}Introduction and Motivation}
Understanding baryon-baryon interactions is critical in nuclear physics because they drive the reaction chain in the atomic nuclei formation. Despite decades of experimental efforts, deuteron is the only dibaryon that has been confirmed. Recent experiments at WASA-at-COSY and SAID amplitude analysis suggest evidence for an unstable light dibaryon, $d^*(2380)$ \cite{WASA-at-COSY:2014dmv}, which needs to be probed further. \refcite{Clement:2016vnl} provides a thorough historical review of the search for dibaryons, covering various experimental efforts to detect other potential dibaryon candidates and theoretical attempts to predict and interpret signatures in the experiments. However, dibaryons in the heavy sector remain less explored, as the prospects of their discovery are relatively low considering the large center-of-mass energies required in their production and the cascade of reaction chains involved in their identification. Despite the lack of discoveries in the heavy sector, it serves as a testing platform to understand baryon-baryon dynamics with relatively less contamination from chiral light quark effects. Recent discoveries in the heavy exotic multi-quark systems at experiments like LHCb, Belle, and ATLAS further motivate studies of dibaryons in the heavy sector. This study employs state-of-the-art lattice QCD techniques to explore dibaryon systems involving heavy quark baryons, with a single flavor at the charm and the strange quark mass. 

Lattice QCD investigations of baryon-baryon scattering and dibaryons are particularly challenging, primarily due to the severe signal-to-noise problem \cite{Lepage:1989hd} that rapidly degrades with time, making it difficult to extract reliable results, especially for systems with a variety of quark line diagrams. Consequently, lattice investigations of dibaryons in the light or strange sector have been conducted at heavy and unphysical pion masses to subside the signal-to-noise ratio problem. For recent lattice studies on dibaryons in the light and strange sectors using various approaches, see \refcite{Iritani:2018vfn, Green:2021qol, Amarasinghe:2021lqa, Aoki:2023qih} and references therein. In the context of heavy quark systems, the large quark mass results in a relatively more precise spectrum due to a better signal-to-noise ratio. Taking advantage of this, the dibaryon with maximum beauty flavor was studied in \refcite{Mathur:2022ovu} and was found to be deeply bound. The spin-3/2 baryons with purely charm or strange quarks ($\Omega_{ccc}$ / $\Omega$) are the lightest excitations in the respective sectors. From a lattice perspective, the $\Omega$ baryon mass serves as a benchmark for scale setting in lattice QCD due to its good signal quality at low computational cost and precise experimental estimate. On the other hand, $\Omega_{ccc}$ baryon is expected to be an ideal candidate for understanding quark confinement dynamics free of any leading light quark effects. Consequently, the two-baryon systems of $\Omega$ and $\Omega_{ccc}$ are particularly suited for understanding baryon-baryon interactions on the lattice.

In this work, we analyze dibaryon systems composed of either two $\Omega_{ccc}$ or two $\Omega$ baryons with total spin $S = 0$ and $S = 2$ and look for signatures of potential bound states in these single-flavored systems. The dibaryon formed by two $\Omega_{ccc}$ baryons is less complex due to relatively suppressed light quark dynamics. A recent study \cite{Lyu:2021qsh} on this dibaryon predicts an attractive potential without Coulomb interactions leading to a bound state, which approaches the unitary regime when the electric charge of the charm quark is switched on. For the dibaryon with six strange quarks, a recent lattice investigation based on the HALQCD procedure suggested weakly attractive baryon interactions \cite{Gongyo:2017fjb}. In contrast, an older study following L\"uscher's finite-volume approach pointed to weakly repulsive interactions \cite{Buchoff:2012ja}. Our investigation of these systems employs a set of baryon-baryon like operators, with maximum overlap with the ground state, in constructing correlation matrices, which are then analyzed variationally to extract the ground state energies. We utilize a set of lattice QCD ensembles to monitor finite size and lattice spacing effects, with the latter being crucial for dibaryons, as emphasized in \refcite{Green:2021qol}. We plan to perform scattering analysis {\it \'a la} L\"uscher in the finite volume as needed, thereby providing a robust investigation of the near-threshold states in these dibaryon systems.

In Section \ref{sec:setup}, we discuss the numerical setup, including the lattice ensembles used and the type of interpolating operators used. The results are discussed in Section \ref{sec:results} followed by discussion and future prospects in Section \ref{sec:conc}.

\section{\label{sec:setup}Numerical Setup}
\begin{figure}[htp]
  \begin{center}
    \includegraphics[width=.66\textwidth]{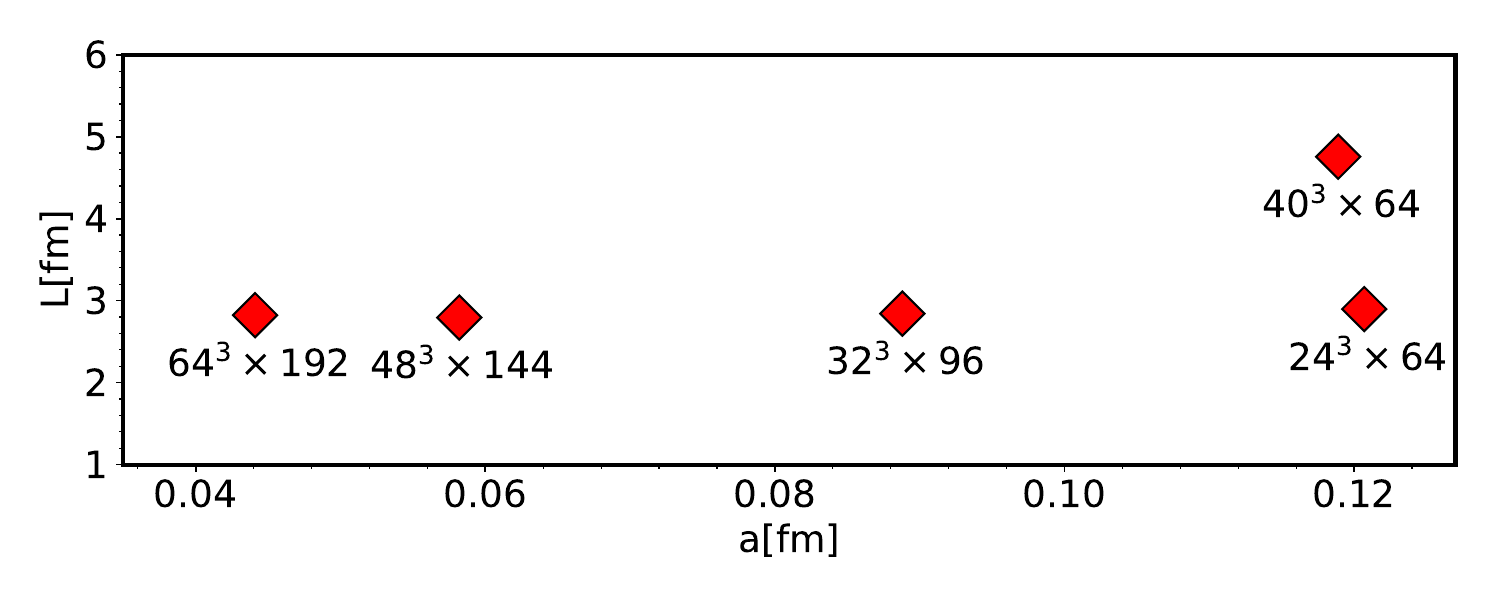}
  \end{center}
  \caption{Overview of the five Lattice QCD ensembles utilized in this study, specified by spatial extent $L$ and lattice spacing $a$. The ensemble $64^3 \times 192$ represents the finest lattice, while the lattice ensembles $40^3 \times 64$ and $24^3 \times 64$ exhibit nearly identical lattice spacings.}
  \label{fig:latt}
\end{figure}

We utilize a set of lattice QCD ensembles with $N_f=2+1+1$ dynamical HISQ fields, two spatial volumes ($L \sim 2.8, 4.7$ fm), and four different lattice spacings, all generated by the MILC collaboration \cite{MILC:2012znn}. The lattice ensemble specifications are depicted in \fig{fig:latt}. We have employed the overlap action \cite{Neuberger:1997fp,Neuberger:1998wv} for valence quark propagators, which have been utilized in our previous works on baryons \cite{Mathur:2018epb, Mathur:2018rwu} and dibaryons \cite{Junnarkar:2019equ, Junnarkar:2022yak}. The details of the charm mass tuning can be found in Refs. \cite{Basak:2012py, Basak:2013oya}. The bare strange quark mass is tuned to the physical point, ensuring the lattice estimate for the hypothetical pseudoscalar $\Bar{s}s$ matches 688.5 MeV \cite{Chakraborty:2014aca}. A wall source and a point sink smearing were used to obtain clean and long ground state plateauing.  We ensure that the results obtained using this asymmetric wall-point setup in evaluating correlation functions are robust when cross-verified using other smearing methods that gradually approach a more sensible symmetric setup \cite{Mathur:2022ovu}. The excited state contamination in the results, assessed from the consistency in ground state plateauing across different smearing procedures, is small compared to the statistical precision. The ground state energies of the baryon and corresponding dibaryons are obtained by fitting eigenvalue correlators, obtained by solving the GEVP, with their expected asymptotic forms. 

For single-flavored (di)baryons, the flavor structure is trivially symmetric, and we utilize only spatially local operators. To make the total wave function of the baryon operator ($\mathcal{O}_b$) anti-symmetric, the spin component remains symmetric since the color structure is naturally anti-symmetric. The only allowed symmetric spin for the $\Omega_{qqq}$ ($q\equiv s,c$) baryon is $S = 3/2$ which is represented by $H^+$ irreducible representation (irrep) for finite cubic lattices \cite{Basak:2005ir}. For a symmetric combination of spinor indices, the $H^+$ irrep has two embeddings \cite{Basak:2005ir} as given in Table \ref{tab:1H_irrep}.

\begin{table}[htbp]
    \centering
    \begin{tabular}{p{0.1\textwidth}||p{0.15\textwidth}|p{0.1\textwidth}||p{0.15\textwidth}|p{0.25\textwidth}}
        $S_z$& Operator [N] & Spin & Operator [R] & Spin \\
        \hline
        +3/2&$^1H_{3/2}$ & \{111\}$_S$&$^2H_{3/2}$ & \{133\}$_S$\\
        +1/2&$^1H_{1/2}$ & \{112\}$_S$&$^2H_{1/2}$ & \{233\}$_S$+\{134\}$_S$+\{143\}$_S$ \\
        -1/2&$^1H_{-1/2}$ & \{122\}$_S$&$^2H_{-1/2}$ & \{144\}$_S$+\{234\}$_S$+\{243\}$_S$\\
        -3/2&$^1H_{-3/2}$ & \{222\}$_S$&$^2H_{-3/2}$ & \{244\}$_S$\\
    \end{tabular}
    \caption{Four rows of baryon operator ($\mathcal{O}_b^a$), where the superscript $a$ refers to the $^1H^+$ and $^2H^+$ irreps. 1,2,3,4 in the columns expressing the spin structure are Dirac spinor indices. 1(3) and 2(4) represent +1/2 and -1/2 spin, respectively, with positive(negative) parity of the four component quark spinor in the Dirac basis. The 1 and 2 components of the quark spinor are referred to as nonrelativistic owing to the velocity independence of their leading components in solutions to the Dirac equation, whereas the 3 and 4 components asymptotes to zero value owing to their nontrivial velocity dependence. Hence, the operator ($^1H^+$) constructed using only 1 and 2 components is referred to as nonrelativistic [N], and $^2H^+$ is referred to as relativistic [R]. Here \{xyz\}$_S=xyz+yzx+zxy$.}
    \label{tab:1H_irrep}
\end{table}

We consider only s-wave interactions in two-baryon systems. The overall dibaryon operator, constructed from two single-baryon operators, must be anti-symmetric under the exchange of the two baryon operators. Since the single-baryon operators are color singlets and are trivially flavor symmetric, the spin structure of a local dibaryon operator must be anti-symmetric, thus limiting the possibility to only even spin operators, i.e., $S = 0, 2$. The dibaryon operator $\mathcal{O}_{d}^{a,b}$ can be constructed from the linear combinations of the single baryon operators ($\mathcal{O}_{b1}^a,\mathcal{O}_{b2}^b$) with the help of CG coefficients as: 
\begin{equation}
    \mathcal{O}_d^{a,b} = \mathcal{O}_{b1}^a. \text{CG}. \mathcal{O}_{b2}^b.
    \label{eqn:CG}
\end{equation}
The interesting cases of $S = 0$ and $2$ are built by projecting continuum-based operators onto the appropriate representations of the octahedral group, following 
\begin{equation}
    ^{[S]}\mathcal{O}^{a,b}_{d,\Lambda,\lambda} = \sum_{S_j}\mathcal{S}_{\Lambda,\lambda}^{S,S_j}\mathcal{O}^{[S,S_j]a,b}_d,
    \label{eqn:subduction}
\end{equation}
where $\mathcal{S}_{\Lambda,\lambda}^{S,S_j}$ are the subduction coefficients, $\Lambda$ is the finite volume irrep and $\lambda$ indicates the row in $\Lambda$. $S = 0$ continuum spin subduces to one-dimensional $A_1^+$ irrep and the five rows of $S = 2$ continuum spin gets distributed over the two-dimensional $E^+$ and the three-dimensional $T_2^+$ irrep. The dibaryon operators for $S = 0, 2$ constructed following this procedure are listed in Eq. \eqref{eqn:ops}. The left superscript $a, b$ on $H$ indicates whether the operator construction is using non-relativistic embedding (N) or relativistic embedding (R); see Table \ref{tab:1H_irrep} for details.
\begin{align}
    ^{[0]}\mathcal{O}^{a,b}_{\text{d,$A_1$,1}} &= \frac{1}{2}\left(~^aH_{3/2}~ ^bH_{-3/2} - ~^aH_{1/2}~ ^bH_{-1/2} + ~^aH_{-1/2}~ ^bH_{1/2} - ~^aH_{-3/2}~ ^bH_{3/2}\right),\nonumber\\
    ^{[2]}\mathcal{O}^{a,b}_{\text{d,$T_2$,1}} &= \frac{1}{\sqrt{2}}\left(~^aH_{3/2}~ ^bH_{-1/2} - ~^aH_{-1/2}~ ^bH_{3/2} \right),\nonumber\\
    ^{[2]}\mathcal{O}^{a,b}_{\text{d,$T_2$,2}} &= \frac{1}{2}\left(~^aH_{3/2}~ ^bH_{1/2} - ~^aH_{1/2}~ ^bH_{3/2} - ~^aH_{-1/2}~ ^bH_{-3/2} + ~^aH_{-3/2}~ ^bH_{-1/2} \right), \label{eqn:ops}\\
    ^{[2]}\mathcal{O}^{a,b}_{\text{d,$T_2$,3}} &= \frac{1}{\sqrt{2}}\left(~^aH_{1/2}~ ^bH_{-3/2} - ~^aH_{-3/2}~ ^bH_{1/2} \right),\nonumber\\
    ^{[2]}\mathcal{O}^{a,b}_{\text{d,$E$,1}} &= \frac{1}{2}\left(~^aH_{3/2}~ ^bH_{-3/2} + ~^aH_{1/2}~ ^bH_{-1/2} -~^aH_{-1/2}~ ^bH_{1/2} -~ ^aH_{-3/2}~ ^bH_{3/2} \right),\nonumber\\
    ^{[2]}\mathcal{O}^{a,b}_{\text{d,$E$,2}} &= \frac{1}{2}\left(~^aH_{3/2}~ ^bH_{1/2} - ~^aH_{1/2}~ ^bH_{3/2} + ~^aH_{-1/2}~ ^bH_{-3/2} - ~^aH_{-3/2}~ ^bH_{-1/2} \right).\nonumber
\end{align}

Given the various possible combinations of embeddings for baryons at the source and sink, we utilized the following operator sets: 
\begin{itemize}
    \item For baryon operators:~~$\{\mathcal{O}^{N}, \mathcal{O}^{R}\}$
    \item For dibaryon Spin 0 operators:~~$\{\mathcal{O}_d^{N,N}, \mathcal{O}_d^{N,R}, \mathcal{O}_d^{R,R}\}$
    \item For dibaryon Spin 2 operators (all five rows of $T_2^+$ and $E^+$):~~$ \{\mathcal{O}_d^{N,R}, \mathcal{O}_d^{R,R}\}$
\end{itemize}
Note that there is no allowed $\mathcal{O}_d^{NN}$ spatially local operator for total spin 2 by the symmetry. With these bases, we could build $2\times2$ correlation matrices for baryons and spin 2 dibaryons, whereas for spin 0 dibaryons, we could construct $3\times3$ correlation matrices. 
\section{\label{sec:results}Results}

\begin{figure}[htp]
  \begin{center}
    \includegraphics[width=.49\textwidth]{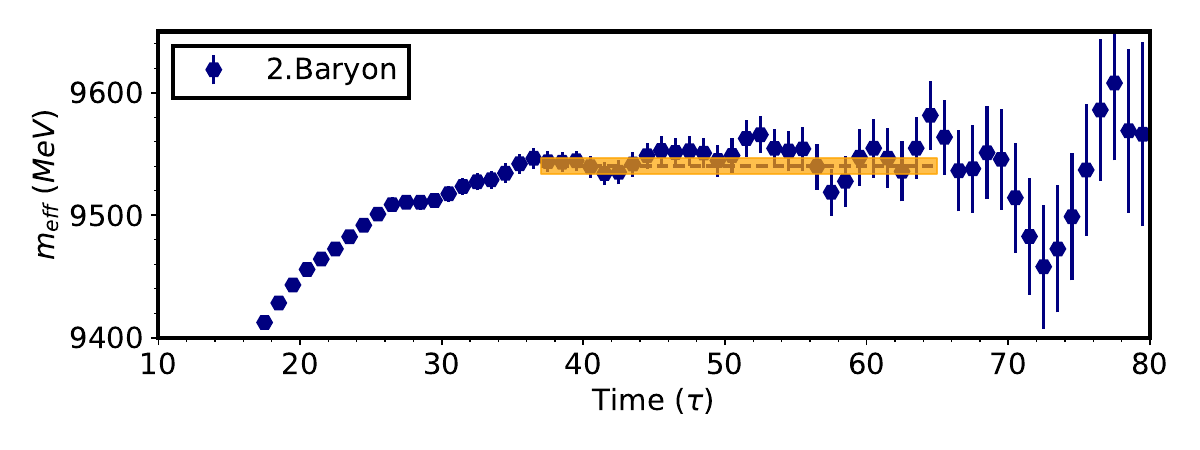}
    \includegraphics[width=.49\textwidth]{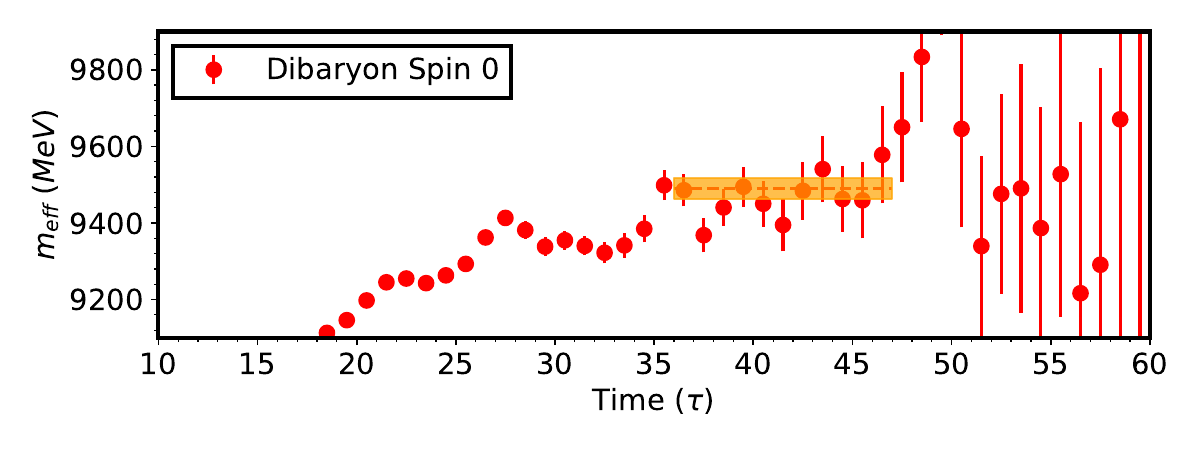}
    \includegraphics[width=.49\textwidth]{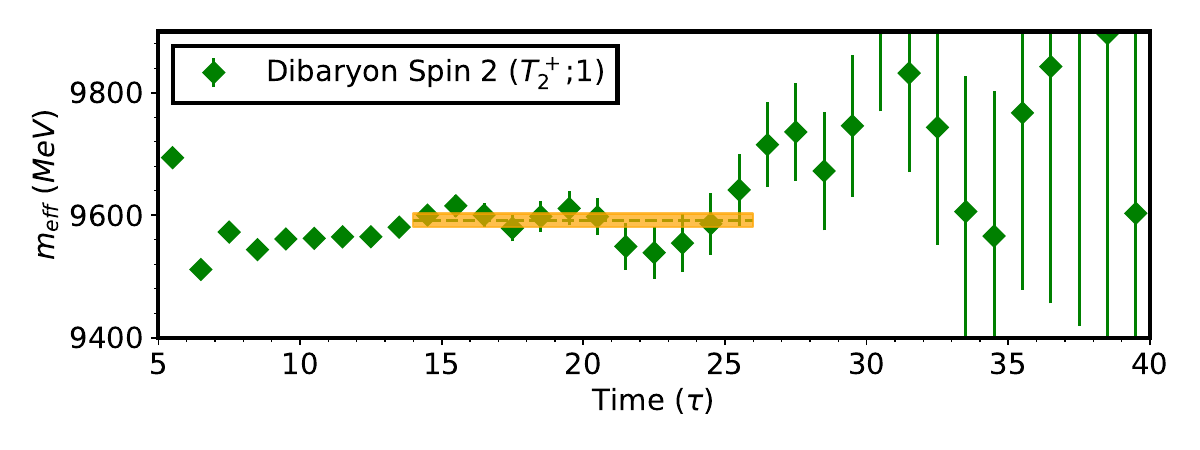}
  \end{center}
  \caption{Effective masses of three systems: two non-interacting baryons (blue), spin 0 dibaryon (red), and spin 2 dibaryon using one row of $T_2^+$ irrep (green). The results are shown for the charm point on $64^3 \times 192$ lattice. The solid bands represent the fit estimates and fitting range used.}
  \label{fig:meff}
\end{figure}

\begin{figure}[htp]
  \begin{center}
    \includegraphics[width=.86\textwidth]{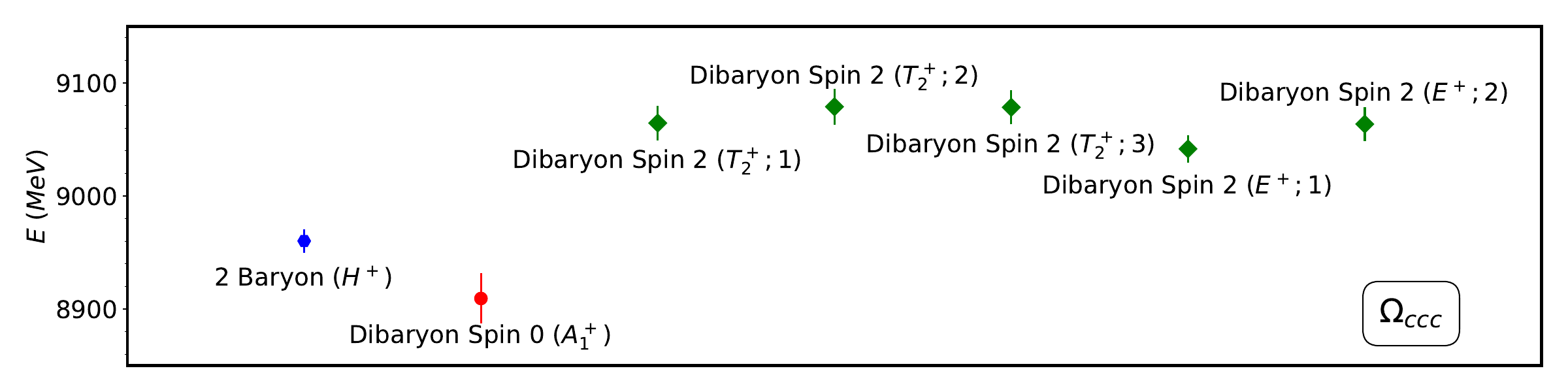}
    \includegraphics[width=.86\textwidth]{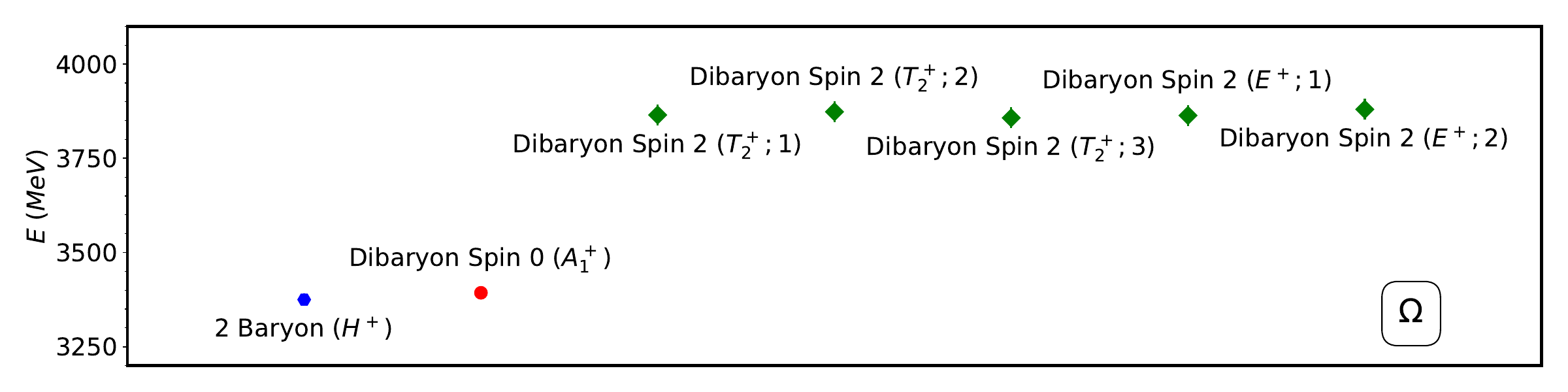}
  \end{center}
  \caption{Fit results for the charm (above) and strange (below) dibaryons on the $48^3 \times 144$ lattice.}
  \label{fig:comp}
\end{figure}

To demonstrate the quality of the signals, we present the effective masses of the ground state energy levels for both baryon and dibaryon ($S = 0$ and $S = 2$) correlation functions, as estimated from the eigenvalue correlators. In \fig{fig:meff}, the results are shown for charm quark mass on the $64^3 \times 192$ lattice across three panels, each representing non-interacting two baryons (top left), spin-0 dibaryon (top right), and spin-2 dibaryon (bottom; using the first row of the dibaryon operator in the $T_2^+$ irrep, see Eq. \eqref{eqn:ops} for reference). The signal for the spin 2 dibaryon is consistent across all five rows of the $E^+$ and $T_2^+$ irreps. Therefore, only one representative case is shown in \fig{fig:meff}. A comprehensive analysis will be presented in our upcoming work. However, for completeness, a comparison plot for all operators is provided in \fig{fig:comp} for the $48^3 \times 144$ lattice, covering both charmed and strange systems. 

It is evident from \fig{fig:meff} that the fitted mass for the spin-2 dibaryon is higher than that of the two non-interacting baryons, while for the spin-0 dibaryon, the fitted value is comparable to the non-interacting case in the scale given. To substantiate this comparison further, we present a compiled plot with fit estimates for different spin systems at the charm and the strange quark masses in \fig{fig:comp}. Clearly, the spin-2 dibaryon ground state is significantly above the non-interacting level. For the strange system, the spin-0 dibaryon ground state energy is statistically consistent with the two-baryon threshold, indicating the absence of a bound state. In contrast, the spin-0 level falls below the two-baryon threshold for the charmed system, suggesting an attractive interaction that can potentially lead to a bound state. 

\begin{figure}[htp]
  \begin{center}
    \includegraphics[width=.8\textwidth]{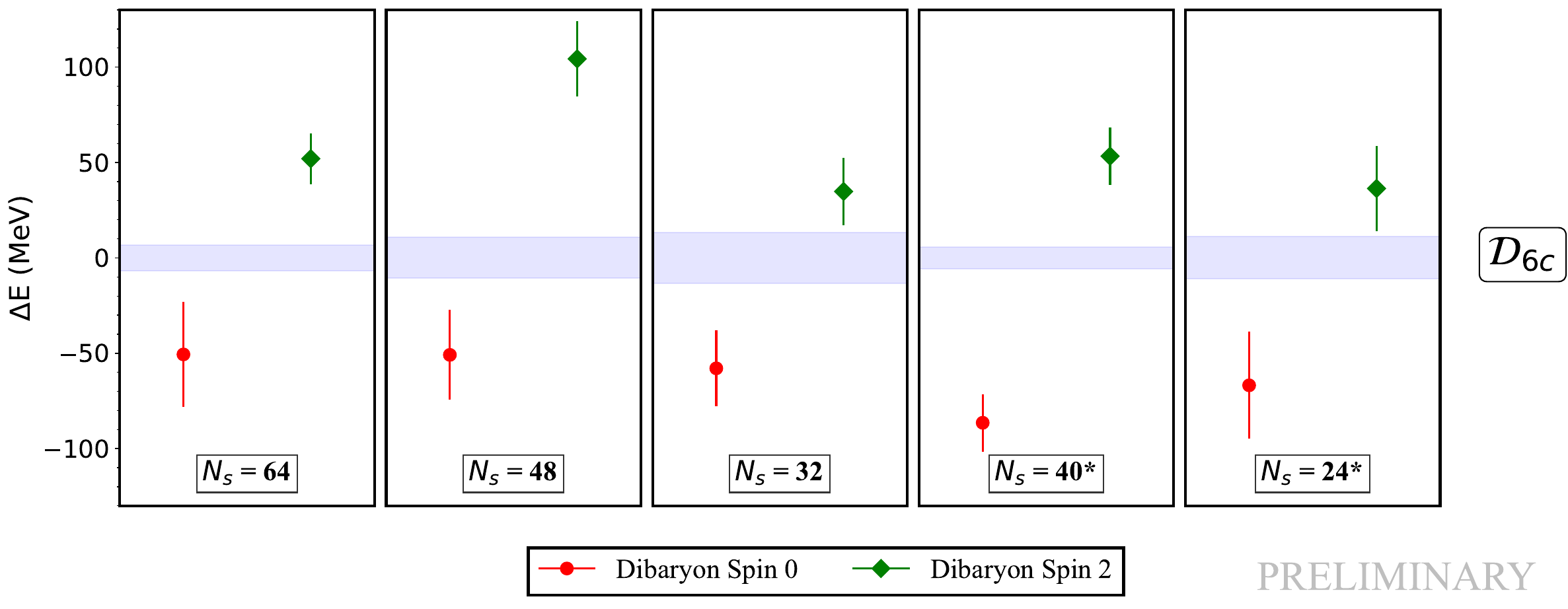}
    \includegraphics[width=.8\textwidth]{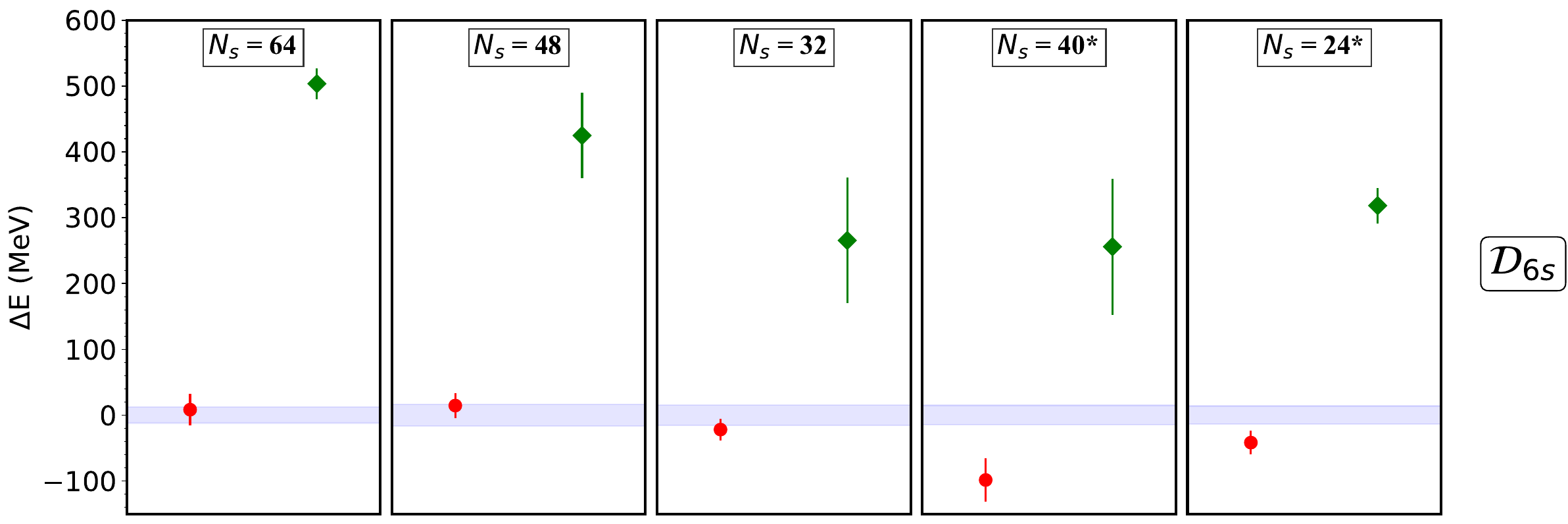}
  \end{center}
  \caption{$\Delta E = m^{fit}_{\text{Dibaryon}} - 2m^{fit}_{\text{Baryon}}$ is shown for both spin-0 and spin-2 dibaryons across all lattice ensembles at the charm (top) and the strange (bottom) quark masses. The solid band represents the non-interacting threshold. For spin-2, only the $T_2^+$ operator row is displayed. The plots are arranged from left to right in order of increasing lattice spacing. The $~^*$ on $N_s = 40$ and $N_s = 24$ lattices indicates that both have nearly the same lattice spacing.}
  \label{fig:main}
\end{figure}

A compiled plot of results from all the ensembles used is shown in \fig{fig:main}. The plot features a solid band representing the non-interacting threshold, against which the spin-0 and spin-2 dibaryon levels are compared through an observable $\Delta E = m^{fit}_{\text{Dibaryon}} - 2m^{fit}_{\text{Baryon}}$. For each case, the embedding combination that provides the best signal is selected from the available options to ensure reliability. The trend indicates that the spin-2 dibaryon consistently lies above the threshold at the strange and the charm quark mass, with this effect becoming more pronounced on finer lattices. The spin-0 strange dibaryon converges to the non-interacting level as the lattice spacing decreases, while the spin-0 charmed dibaryon is slightly below the threshold, suggesting a possible bound state. 

\section{\label{sec:conc}Outlook and next steps}
We investigated the dibaryon systems with single flavor quarks in strange and charm sectors, utilizing several lattice ensembles with multiple lattice spacings. This setup, with an extended operator basis and a wall-source to point-sink smearing procedure, provided clean signals with good overlap to the ground state, enhancing the reliability of our results. In the $S = 2$ channel, we observed a positive shift, indicative of a repulsive interaction suggesting the absence of bound states in both strange and charm systems. In the charm sector, the spin-zero results show a slight tendency toward negative shifts, consistent across all lattices but of small magnitude, prompting further investigation. Conversely, the spin-zero results broadly suggest a non-interacting scenario in the strange sector, potentially indicating no bound states. More definitive conclusions can be drawn with a detailed finite-volume amplitude study which is underway. The insights and experience we gain from these studies provide a platform for lattice investigations of other dibaryon systems, such as that of the $d^*(2380)$ resonance.

\vspace{20 pt} 
\noindent \textsc{Acknowledgments:}~ 
This work is supported by the Department of Atomic Energy, Government of India, under Project Identification Number RTI 4002. NSD would like to thank IMSc for the travel grant, which enabled him to present this work at the Lattice conference.  MP gratefully acknowledges support from the Department of Science and Technology, India, SERB Start-up Research Grant No. SRG/2023/001235. We are thankful to the MILC collaboration and, in particular, to S. Gottlieb for providing us with the HISQ lattice ensembles. Computations were carried out on the Cray-XC30 of ILGTI, TIFR, and the computing clusters at DTP, TIFR Mumbai, and IMSc Chennai.


\bibliographystyle{JHEP}
\bibliography{proc}
\end{document}